\documentclass{article}
\usepackage{graphicx}
\usepackage[a4paper, total={6.5in, 8.5in}]{geometry}
\usepackage[numbers,sort&compress]{natbib}
\usepackage[switch]{lineno}
\usepackage{titlesec}
\usepackage{amssymb}
\usepackage{authblk}
\titleformat*{\section}{\centering\large\bfseries}
\titleformat*{\subsection}{\centering\small\bfseries}
\begin{document}
\hyphenation{mo-le-cules}
\hyphenation{project-ruimte}
\sloppy

\title{Deceleration of a supersonic beam of SrF molecules to 120~m/s}

\date{}
\author{S. C. Mathavan}
\author{A. Zapara}
\author{Q. Esajas}
\author{S. Hoekstra}
\affil{Van Swinderen Institute, University of Groningen, Zernikelaan 25, 9747 AA, Groningen, The Netherlands}

\twocolumn[
 \begin{@twocolumnfalse}
    \maketitle
    \begin{abstract}
   We report on the deceleration of a beam of SrF molecules from 290 to 120~m/s. Following supersonic expansion, the molecules in the $X^2\Sigma$ ($v=0$, $N=1$) low-field seeking states are trapped by the moving potential wells of a traveling-wave Stark decelerator. With a deceleration strength of 9.6 km/s$^2$ we have demonstrated the removal of 85\% of the initial kinetic energy in a 4 meter long modular decelerator. The absolute amount of kinetic energy removed is a factor 1.5 higher compared to previous Stark deceleration experiments. The demonstrated decelerator provides a novel tool for the creation of highly collimated and slow beams of heavy diatomic molecules, which serve as a good starting point for high-precision tests of fundamental physics. \bigskip
    \end{abstract}
    \end{@twocolumnfalse}
    ]

\section{Introduction}
The creation of slow beams of heavy diatomic molecules has been a long-standing goal, motivated by the application of such molecules in the measurement of the electron electric dipole moment and tests of parity violation~\cite{2014Sci...343..269B, Hudson:2011hs,Demille:2008he,DeMille:2015hl}. The long interaction time offered by slow beams is attractive as the sensitivity of the measurement is linearly improved by it.

All neutral molecules proposed and used for tests of fundamental symmetries so far are composed of at least one atom that can not be laser-cooled. This restricts the methods that can be used to direct cooling, where the molecule of choice is usually produced and cooled in a molecular beam and then decelerated and prepared for the measurement using a combination of techniques. A number of such direct cooling methods have been applied to this purpose until now. Cryogenic beam sources have made great progress in recent years, producing beams of molecules with a forward velocity in the range of 200~m/s down to 65~m/s~\cite{Lu:2011dz,Hutzler:2012kd,Patterson:2015in}. The transverse velocity spread of such beams is typically around 30-50~m/s, leading to a rapid decrease of beam density with distance from the source.

Another area where significant advances have recently been made is the laser cooling of molecular beams of SrF~\cite{2010Natur.467..820S,Barry:2011ur}, CaF~\cite{2014PhRvA..89e3416Z,2016arXiv160302787H}, YO~\cite{2013PhRvL.110n3001H} and SrOH~\cite{2016JPhB...49m4002K}. Even though the same challenge of transverse spreading of the beam once it is slow~\cite{DeMille:2013jq} also holds for this approach, nonetheless slow SrF molecules have been collected and trapped in a magneto-optical trap~\cite{2016PhRvL.116f3004N}.

In the technique of Stark deceleration, electric fields are used to exert a force on polar molecules. This technique has been used successfully to decelerate and trap a range of molecules~\cite{Meerakker:2008ii, Meerakker:2012ChemRev}, mostly with a mass below 40 atomic mass units. The challenge in the Stark deceleration of heavier molecules is twofold. First of all their mass leads to a large initial kinetic energy in the supersonic expansion, requiring a longer decelerator. Secondly, especially for the interesting class of alkaline-earth monofluorides, their rotational energy level structure leads to an unfavorable Stark shift. As a result, the molecules will be lost from the deceleration process if the electric fields are too high. This limitation motivated the approach to decelerate such molecules in their high-field seeking ground state~\cite{2002PhRvL..88m3003B,Tarbutt:2004du,2006JPhB...39R.263B,2008PhRvA..77c1404W,PhysRevA.78.033421}. Although demonstrated to work in principle, this approach suffered from low acceptance and instabilities due to mechanical misalignment~\cite{Tarbutt:2008gl}.

An alternative approach is traveling-wave Stark deceleration, in which packets of molecules are confined by a traveling potential well, formed by electric field gradients that exert a force on polar molecules. What sets traveling-wave Stark deceleration apart from the previously mentioned approaches is that the molecules remain transversely confined throughout the deceleration~\cite{2010PhRvA..81e1401O,vandenBerg:2012uf,vandenBerg201422,QuinteroPerez:2013bk,2013PhRvA..88d3424J}. The resulting beam has a narrow velocity spread, both longitudinally and transversely, of just a few m/s. This means that a large fraction of the molecules that fall within the acceptance of the Stark decelerator can be used for the experiment. Compared to the other methods, the largest reduction in the number of molecules is therefore moved to the high velocities in the beginning of the Stark deceleration process.

Traveling-wave deceleration has so far been demonstrated on CO~\cite{2010PhRvA..81e1401O}, SrF~\cite{vandenBerg201422}, NH$_3$~\cite{2013PhRvA..88d3424J,QuinteroPerez:2013bk}, CH$_3$F~\cite{2015PhRvA..92b3404M} and YbF~\cite{2012PhRvA..86b1404B}, but of these only NH$_3$ and CH$_3$F molecules have been decelerated to a standstill and thereby electrically trapped. Since for the creation of trapped samples or slow beams of heavier ($>100$ amu) molecules, such as SrF, BaF and YbF, much more kinetic energy has to be removed, we set out to build a long and modular traveling-wave decelerator. We have reported on the first operation of a 2~meter long decelerator consisting of 4 modules previously~\cite{vandenBerg201422}. Here we report on the successful operation of 8 modules of this decelerator, with a total length of 4 meter, and demonstrate the removal of 85\% of the kinetic energy of a 290~m/s supersonic SrF beam. The operation of this longer decelerator is an important test of its stability and of the high-voltage electronics. In following experiments a final module, adapted to provide good optical access, will be added to the apparatus to bring the molecules to a standstill. The slowest SrF molecules we have produced until now are traveling at a velocity of $120 \pm 2$ m/s with a transverse velocity spread of 3.5~m/s.

\section{Experimental methods}
\subsubsection*{Making the SrF}
\begin{figure*}
	\includegraphics[width=\textwidth]{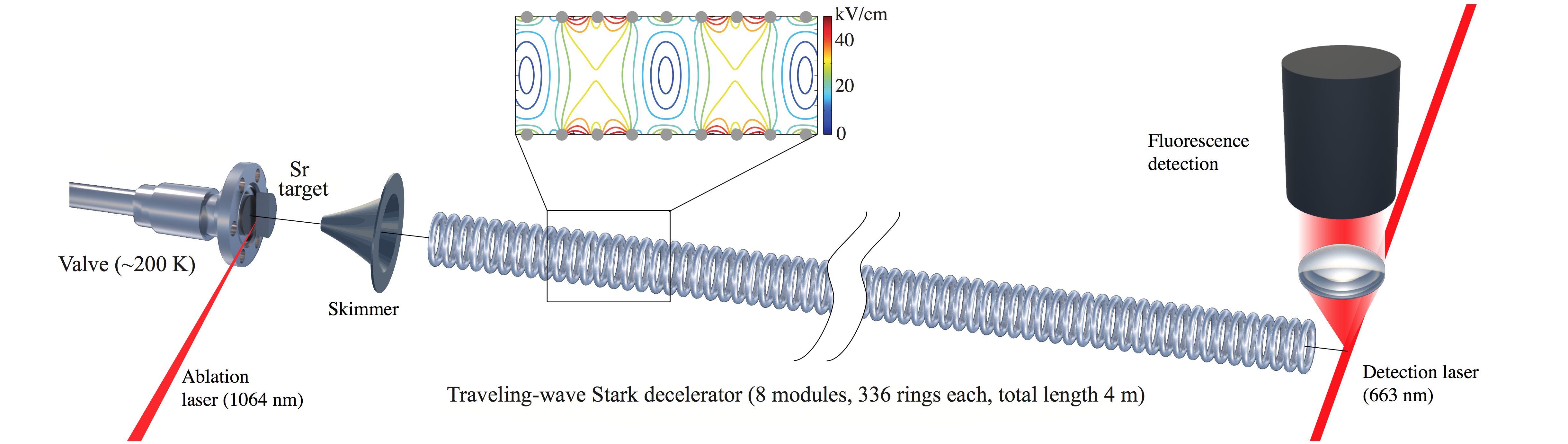}
	\caption{\label{schematic}Schematic overview of the experimental setup. SrF molecules are created by laser ablation from a Sr metal target in the presence of SF$ _6 $ gas. Following pulsed supersonic expansion through a skimmer a fraction of the molecules is captured by the traveling potential of the decelerator. Molecules reaching the end of the decelerator are detected by laser-induced fluorescence.}
\end{figure*}

We create SrF molecules by laser ablating a Sr metal target in the presence of SF$_6$ gas. The target is placed inside a compact source chamber which is pumped to a base pressure of ~$2\times 10^{-8} $ mbar. Inside the source chamber a valve is used to create a pulsed supersonic beam from a mixture of Xe and 3 \% SF$_6$ gas at a repetition rate of 10 Hz. During operation the source pressure increases to $2\times 10^{-4} $ mbar. We use laser pulses from a Q-switched Nd:YAG laser at 1064 nm with a pulse energy of 60 mJ and an elliptical beam spot of $3\times1$ mm to ablate the Sr metal target. The target is mounted on a piezo-electric translation stage, which is moved during the ablation process after every few shots. For cooling with cold Nitrogen gas, a copper housing with a gas flow channel encloses the valve. During the deceleration measurements reported here, the valve was cooled to a temperature of $\sim$ 200~K, resulting in an average molecular beam speed of 315~m/s. This velocity depends on our control of the temperature and the backing pressure, since we operate close to the condensation point of the Xenon gas. In a pure Xenon expansion with optimal conditions we have previously achieved a velocity centered around 280~m/s~\cite{Berg:2014tt}. The central part of the beam passes through a skimmer with exit diameter of 2~mm, placed 60~mm downstream from the ablation spot, before entering the decelerator. A schematic overview of the experimental setup is shown in Fig.~\ref{schematic}.

\subsubsection*{Traveling-wave decelerator}
We decelerate $^{88}$SrF molecules in the (1,0) rotational state. The numbers in the brackets are quantum numbers \textit{N} and \textit{M$_N$}, where \textit{N} is the rotational quantum number in the zero field and \textit{M$_N$} the projection of \textit{N} on the electric field axis. This state is sensitive for parity violation measurements~\cite{Demille:2008he} and it is the one which has already been shown to be laser cooled~\cite{2010Natur.467..820S}. The decelerator consists of modules of 50~cm length, allowing us to build a decelerator of the required length. The 4~m decelerator used for the measurement reported here is constructed by connecting eight modules. For further details of the design of our modular traveling-wave decelerator we refer to~\cite{vandenBerg:2012uf}.

Deceleration of molecules in a traveling-wave decelerator is achieved by trapping the molecules in a moving electric field created by oscillating voltages of the form $V_n(t)=V_0 \sin(2\pi ft+\frac{n\pi}{4})$. These voltages are applied to 8 sets of ring electrodes, and deceleration is obtained by sweeping the frequency \textit{f} of these voltages to lower values. We use arbitrary waveform generators to create the required voltage waveform which is then amplified by 8 custom high-voltage amplifiers that are capable of maintaining the sinusoidal wave with an amplitude of 5 kV on the capacitive load of the decelerator. We measured the capacitance of the 4 m decelerator and found that within 10\% accuracy each of the sets of electrodes has a total capacitance of 200~pF to its nearest neighbours, which is twice the value of the capacitance with the 2~m decelerator. The peak-to-peak amplitude of 10~kV is optimal for the low-field seeking SrF(1,0) state; for efficient deceleration of the next rotational SrF(2,0) state a higher amplitude would be required. SrF has a turning point in the Stark curve of the $(N,M_N)=(1,0)$ state at a field strength of 25~kV/cm, limiting the depth of the traveling potential well created by the electric fields to 0.16~cm$^{-1}$. 

\subsubsection*{Detection}
We use the strong $ A^{2} \Pi_{1/2}(v=0,J=1/2)$ $\leftarrow$ $X^{2}\Sigma^{+}$ $(v=0,N=1)$ transition at 663~nm for resonant fluorescence detection. The $N=1$ rotational level of $^{88}$SrF is split into four components due to spin-rotation coupling and hyperfine structure. We address these four states using sidebands that are created by an electro-optic modulator.

The continuous wave detection laser beam with a $1/e$ diameter of 3~mm and power of 2~mW crosses the molecular beam after the decelerator. The fluorescent light is collected by a plano-convex lens and a mirror, and passed through a bandpass interference filter centered around 661~nm. After the filter, a lens focuses this light through an adjustable iris onto the cathode of a photomultiplier tube (PMT). The data acquisition system records the arrival times of the photons, resulting in binned time-of-flight profiles that are presented in the next section.

\section{Results}
\subsubsection*{Time-of-flight profiles}
\begin{figure*}
	\includegraphics[width=\textwidth]{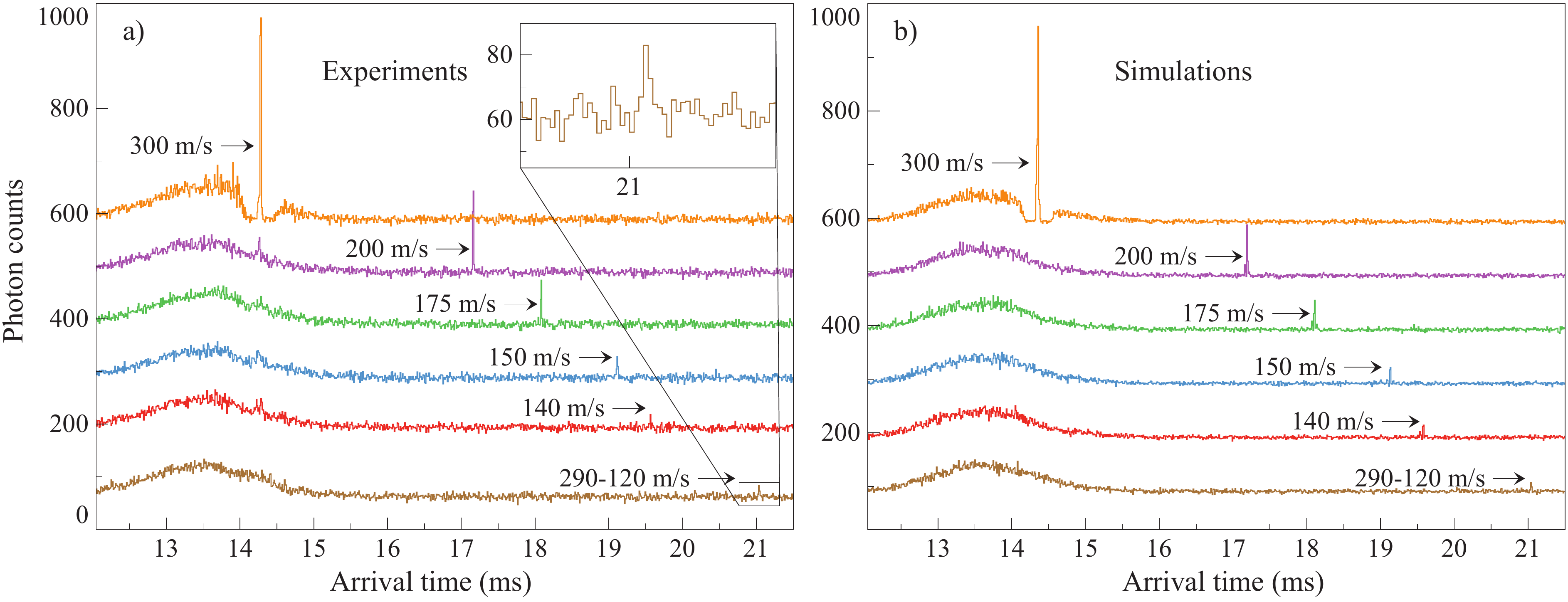}
	\caption{\label{results}Experimental a) and simulated b) time of flight profiles (binsize 10 $\mu$s) showing the deceleration of SrF molecules from a starting velocity of 300 m/s (top 5 curves) and 290 m/s (bottom curve). The final velocities are indicated by the labels. A vertical offset is added for clarity.}
\end{figure*}

Fig. \ref{results}a) shows the deceleration results of SrF molecules in their $(N,M_N)=(1,0)$ rotational state in terms of time-of-flight profiles with a binsize of 10~$\mu$s. Each plot in the figure represents a measurement of 10~minutes. The laser ablation of the pill corresponds to $t=0$. For clarity we have added a vertical offset to the histograms. The uppermost histogram shows the arrival time of molecules in a guiding mode with a constant velocity of 300~m/s. The central peak around 14.3~ms is formed by molecules that are within the longitudinal phase-space acceptance of the decelerator. The two adjacent wings correspond to molecules that are outside the longitudinal but inside the transverse acceptance. The guiding velocity is slightly lower than the mean velocity of the initial molecular beam. The next four histograms demonstrate deceleration results from 300~m/s with increasing deceleration strengths, for which the final velocity is indicated. The delayed arrival of the decelerated molecules is accompanied by a decrease of the number of molecules, due to the corresponding reduction of the volume of the phase-space stability. The bottom curve shows the deceleration from 290~m/s to 120~m/s, which was done with the coldest valve ($\sim$190~K). Due to clogging of the valve we did not obtain a full range of deceleration strengths for this initial velocity. The last result demonstrates the operation of the 4~meter long decelerator with a constant deceleration of 9.6~km/s$^2$, which for these initial conditions corresponds to the removal of 85\% of the initial kinetic energy.

\subsubsection*{Trajectory simulations}
In order to analyze the experimental results we have performed trajectory simulations of the deceleration process, resulting in the simulated time-of-flight profiles that are displayed in Fig.~\ref{results}b). Care has been taken to ensure the simulations are numerically stable. The beam conditions of the simulations have been matched to those of the experiment. The trajectory simulations reproduce all essential features of the measured time-of-flight profiles, including the decrease of the guided peaks and the relative intensity of the non-decelerated part. In this case it is possible to reconstruct the kinematic properties of the molecular beam at any stage of the deceleration. We can derive that the mean longitudinal velocity of the SrF beam after the source chamber is 315~m/s with a full width at half maximum (FWHM) of 40~m/s. We also deduce from the simulations that the final transverse velocity distribution of the decelerated packets of SrF molecules is well described by a Gaussian with a FWHM of 3.3~m/s. The longitudinal velocity spread depends on the deceleration strength and ranges from a FWHM of 6~m/s at 300~m/s down to only 1.7~m/s at a forward velocity of 140~m/s. There are some slight but noticeable differences between simulations and experimental results that can be attributed to the systematic effects and mechanisms that are not included in the simulation code, which are: valve temperature stability during the measurements, imperfections of the waveforms, possible non-adiabatic losses or parametric heating mechanisms. However, under given conditions all of them play minor role for the deceleration efficiency.

\section{Discussion}
\subsubsection*{Decelerator performance}
The overall performance of the decelerator can be quantitatively described by the deceleration efficiency, which we define as the fraction of molecules that can be decelerated at a given deceleration strength compared to the total amount of molecules that can be guided at the initial velocity. This number can be found from the experimental results as well as from the trajectory simulations. Since the longitudinal acceptance of the decelerator goes down with increasing deceleration strength, the fraction of trapped molecules also decreases. In case of deceleration of SrF molecules from 290 to 120~m/s the aforementioned efficiency is about 6-7\% from the experimental results which under these experimental conditions is in good agreement with the simulations. The demonstrated deceleration efficiency is also consistent with the previous results~\cite{vandenBerg201422} with the 2~meter long decelerator. We have however identified loss mechanisms under other circumstances, mainly at higher initial velocities, which we will report on in a future publication.

\subsubsection*{Number of molecules}
To obtain the number of decelerated molecules from the number of detected photons, we determined the detection efficiency. We estimate this detection efficiency to be $ (0.25 \pm 0.05)\%$ taking into consideration the following factors: collection solid angle, quantum efficiency of the PMT, longitudinal and transverse velocity spread of the beam, average number of scattered photons/molecule (estimated at 3.5) and the transmission of the optics. With this efficiency the total number of molecules per shot reaching the end of the decelerator in the SrF(1,0) state is $(5.6 \pm 1) \times 10^3$. These arrive within a time window of 10 ms around the guided peak. For the measurements with an initial velocity of 300~m/s, the detected signal in a time window of 100 $\mu$s around the guided peak corresponds to $440 \pm 90$~molecules/shot, which decreases to $28 \pm 6$ molecules/shot in the measurement with the maximum deceleration strength (300-140~m/s). When selecting 290~m/s as the initial velocity, we obtain $230 \pm 50$~molecules/shot in the guided peak (data not shown) and $16 \pm 3$ molecules/shot in the decelerated peak with final velocity of 120~m/s.

We would like to point out that the low number of molecules per shot in the deceleration experiments is not caused by the performance of the decelerator but by the performance of the SrF supersonic source. Compared to our source, a previously reported beam of similarly produced YbF molecules~\cite{Tarbutt:2002wv} is about 1 order of magnitude more intense. 

\subsubsection*{Possibilities with other molecules}
\begin{figure}[ht]
	\includegraphics[width=0.5\textwidth]{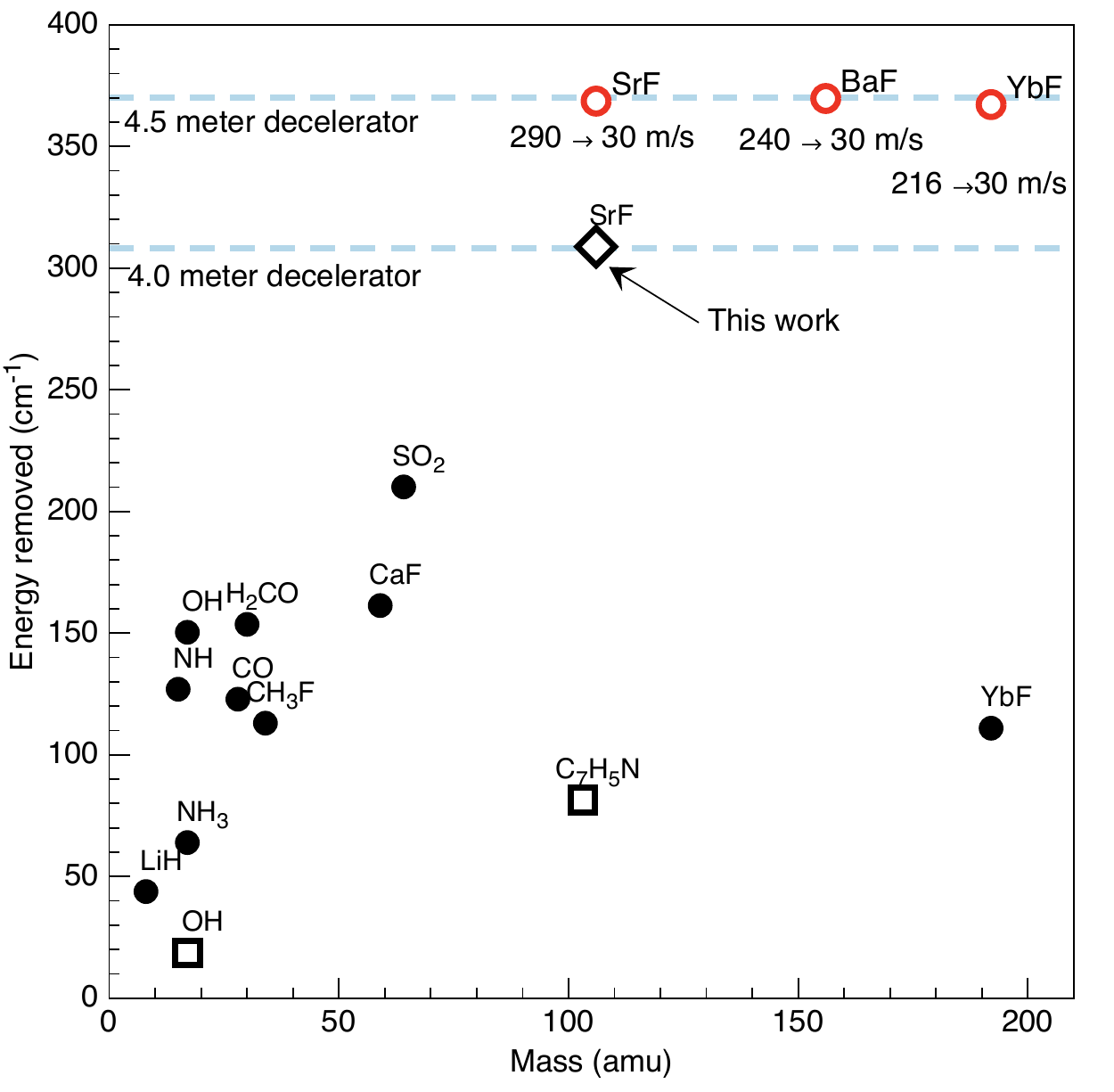}
	\caption{\label{figure3}The total kinetic energy removed in the deceleration reported in this work (diamond) compared to previous Stark deceleration experiments with low-field seeking molecules (filled circles) and high-field seeking molecules (open squares). The deceleration possible with a decelerator of 4.5 m length is also indicated (red open circles).}
\end{figure}

To put the reported results on the deceleration of SrF into perspective, we have plotted in Fig.~\ref{figure3} the total amount of kinetic energy removed, together with an analysis of previous Stark deceleration experiments~\cite{Bucicov:2007wc,Hoekstra:2007vxa,2007JChPh.127v1102G,2008PhRvA..77c1404W,Anonymous:30qkyq7f,Meerakker:2005vh,2012PhRvA..86b1404B,2006PhRvA..73f3404H,2015PhRvA..92b3404M,1999PhRvL..83.1558B,PhysRevA.78.033421}. For the vertical energy axis we have chosen units of wavenumber (cm$^{-1}$), on the horizontal axis the molecular mass is indicated. As can be expected from the length of our decelerator, the amount of energy removed is larger than any previously reported values, by about a factor 1.5. The longest Stark decelerator operated previously was 2.6 meters long~\cite{2009PhRvA..79b3410S}.

By adding one additional module to the decelerator we can decelerate SrF molecules from the cooled Xe supersonic expansion at 290~m/s to standstill. To decelerate even heavier molecules with similar Stark shifts, such as BaF and YbF, we have to decrease the initial velocity further. This could be done for example by using a cryogenic buffer gas source, operated at 20~K with Neon gas, close to the supersonic regime. Typical velocities of 180 to 200~m/s have been demonstrated~\cite{Hutzler:2012kd}, which are well below the initial velocities that are required to decelerate to a slow beam of 30~m/s (or standstill) as indicated by the red points in Fig.~\ref{figure3}. To compensate for their different masses the initial velocity that can be decelerated with 7\% efficiency is also indicated in the figure (open circles). Especially when operated in the fast beam regime, the cryogenic sources can also deliver a much higher phase-space density per rotational state compared to the supersonic beam, offering exciting prospects for the creation of intense and collimated slow beams of heavy diatomic molecules.

Besides the possibilities with heavier molecules, lighter molecules in states with a Stark shift that is smaller than that of SrF(1,0) can still be decelerated using the long decelerator demonstrated here. 

\section{Conclusion}
We report on the deceleration of a supersonic beam of SrF molecules to a final velocity of 120~m/s, which corresponds to the removal of 85\% of the initial kinetic energy. These experiments demonstrate the successful operation of a 4~meter long traveling-wave Stark decelerator, removing 1.5 times more kinetic energy than any previous experiment. This is a novel tool for deceleration of heavy diatomic molecules to arbitrarily slow beams with a very small velocity spread. The number of molecules can be increased by improving the source intensity, and the efficiency of the deceleration process can be increased by decreasing the initial velocity of the beam. By combining the traveling-wave Stark decelerator with a cryogenic source slow, intense and highly collimated beams of heavy diatomic molecules can be produced, which serve as an excellent starting point for future precision tests of fundamental physics.

\section{Acknowledgment}
We acknowledge the expert technical assistance of Leo Huisman, Imko Smid and
the KVI mechanical workshop, and Corine Meinema for her contributions in the early stages of this project. This work is part of the research programme of the Foundation for Fundamental Research on Matter (FOM), which is part of the Netherlands Organization for Scientific Research (NWO) (FOM Programs nr. 114 and 125, Projectruimte 11PR2858, VIDI 680-47-519).

\bibliographystyle{unsrt}

\end{document}